\documentclass[preprint,aps,superscriptaddress,nofootinbib,tightenlines]{revtex4}

\usepackage{epsfig}

\def\OMIT#1{{}}

\def\si{^1 \hskip -0.03in S _0}
\def\siii{^3 \hskip -0.025in S _1}
\def\diii{^3 \hskip -0.03in D _1}
\def\pislash{ {\pi\hskip-0.54em /} }

\def\nopi{ {\rm EFT}(\pislash) }
\newcommand{\gsim}{\raisebox{-0.7ex}{$\stackrel{\textstyle >}{\sim}$ }}
\newcommand{\lsim}{\raisebox{-0.7ex}{$\stackrel{\textstyle <}{\sim}$ }}

\begin{document}

\preprint{\vbox{
\hbox{UNH-03-02}
\hbox{LBNL-54092}
\hbox{NT@UW-03-033}
}}

\vphantom{}

\title{Two Nucleons on a Lattice}

\author{ S.R.~Beane}
\affiliation{Department of Physics, University of New Hampshire,
Durham, NH 03824-3568.}
\affiliation{Jefferson Laboratory, 12000 Jefferson Avenue, 
Newport News, VA 23606.}
\author{P.F.~Bedaque}
\affiliation{Lawrence-Berkeley Laboratory, Berkeley,
CA 94720.}
\author{A.~Parre{\~n}o}
\affiliation{Dept. ECM, Facultat de F\'{\i}sica, Universitat de Barcelona,
E-08028, Barcelona, Spain.}
\author{M.J.~Savage}
\affiliation{Department of Physics, University of Washington, 
Seattle, WA 98195-1560.\\
\qquad}

\vphantom{}
\vskip 0.5cm
\begin{abstract} 
\vskip 0.5cm
\noindent 

\noindent 
The two-nucleon sector is near an infrared fixed point of QCD and as a
result the S-wave scattering lengths are unnaturally large compared to
the effective ranges and shape parameters.  It is usually assumed that
a lattice QCD simulation of the two-nucleon sector will require a
lattice that is much larger than the scattering lengths in order to
extract quantitative information.  In this paper we point out that
this does not have to be the case: lattice QCD simulations on much smaller
lattices will produce rigorous results for nuclear physics.
\end{abstract}

\maketitle

\vfill\eject

\section{Introduction}

\noindent One of the central goals of nuclear physics is to make
rigorous predictions for both elastic and inelastic processes in
multi-nucleon systems directly from QCD.  The only presently-available
technique to achieve this goal is lattice QCD, where space-time is
discretized and QCD Green functions are evaluated in Euclidean space.
Unfortunately, at present, the variety of processes that can be
addressed with lattice QCD is quite limited.  The currently-available
computational power restricts not only the sizes of lattices that can
be utilized, but also the lattice spacings and quark masses that can
be simulated.  Moreover, the Maiani-Testa theorem~\cite{Maiani:ca}
precludes determination of scattering amplitudes away from kinematic
thresholds from Euclidean-space Green functions at infinite volume.
However, by generalizing a result from non-relativistic quantum
mechanics~\cite{yang} to quantum field theory,
L{\"u}scher~\cite{Luscher:1986pf,Luscher:1990ux} realized that one can
access $2\rightarrow 2$ scattering amplitudes from lattice simulations
performed at finite volume.  Significant progress has been made using
this finite-volume technique to determine the low-energy $\pi\pi$
phase shifts directly from QCD, e.g. Ref.~\cite{Aoki:2002ny}.
However, only one lattice QCD calculation of the nucleon-nucleon (NN)
scattering lengths~\cite{Fukugita:1994ve} has been attempted, and it
was a quenched simulation with heavy pions~\footnote{For a
recent review of hadron-hadron interactions on the lattice, see
Ref.~\cite{Fiebig:2002kg}.}.

When contemplating computing nuclear observables with lattice QCD one
naively assumes that the lattice must be much larger than the systems
being simulated, so that the systems on the lattice resemble those at
infinite-volume.  This would mean, for instance, that when computing
the rate for the simplest inelastic nuclear process, $np\rightarrow
d\gamma$, which near threshold involves radiative capture from the
$\si$ channel, a lattice of size $L\gg | a^{(\si)}|, \ | a^{(\siii)}|$
is required, where $a^{(\si)}$ and $a^{(\siii)}$ are the $\si$ and $\siii$
NN scattering lengths, respectively. Given that $a^{(\si)} = -23.714~{\rm fm}$, such a
calculation would have to await a future in which computational power
is sufficient to handle volumes of this size. Fortunately, as we will
see, this argument is not correct. 

There is a sizable separation of length scales in nuclear physics, due
to the fact that nature has chosen to be very near an infrared fixed
point of QCD~\cite{Kaplan:1998tg,Kaplan:1998we,Birse}.  As a result,
the scattering lengths in both $S-$wave channels are unnaturally-large
compared to all typical strong-interaction length scales, including
the range of the nuclear potential which is determined by the pion
Compton wavelength.  Perhaps counter-intuitively, in simulating
two-nucleon processes, the relevant lengths scales are those of the
nuclear potential and {\it not} the scattering lengths, and thus as
long as the lattice is large compared to the inverse of the pion mass
one can in principle ``simply '' determine matrix elements and
scattering parameters.  Furthermore, quantitative information about the
two-nucleon sector can be extracted from simulations on even smaller
lattices.  However, the theoretical analysis that would be required
for such an extraction is significantly more complex, and in this work
we restrict ourselves to lattices that are much larger than the pion
Compton wavelength~\footnote{There is a second technique that can be
used to extract information about nuclear processes from small
lattices.  If simulations are performed on lattices with quark masses
that are somewhat different from their physical values, the scattering
lengths will, most likely, no longer be unnaturally
large~\cite{Beane:2002vq,Beane:2002xf,Epelbaum:2002gb}.  By using the
pionful effective field theory to calculate the quark mass dependence
of the scattering parameters, the results of such simulations can be
related to those at the physical values of the quark masses.  The
expansion parameters in the pionful theory are not exceptionally
small, and so higher order calculations will need to be performed in
order to have reliable extrapolations.  We do not explore this option
in this work.  }.

Perhaps the motivation for a lattice calculation of the
radiative-capture process $np\rightarrow d\gamma$ is less than
compelling as the cross-section and contributing multipoles at
low-energies are well-measured.  High-precision data recently
collected~\cite{Tornow:2003ze} with HI$\gamma$S agrees well with
calculations~\cite{Chen:1999bg,Rupak:1999rk} in the pionless effective
field theory~\cite{Chen:1999tn,vanKolck:1998bw}, $\nopi$, and also
with the best modern potential models (see Ref.~\cite{Tornow:2003ze}).
However, weak processes such as $\nu d\rightarrow \nu n p$, which play
a central role in the determination of solar-neutrino fluxes from the
sun, depend upon two-body weak
currents~\cite{Butler:1999sv,Butler:2000zp} that presently are
determined with significant uncertainties from reactor
experiments~\cite{Butler:2002cw} and are also determined from the
$\beta$-decay of tritium~\cite{Park:2002yp} with unknown systematic
uncertainties.  Recently they have been determined by 
the Sudbury Neutrino Observatory (SNO) from a fit
to the neutrino fluxes~\cite{Chen:2002pv}, but again with large
uncertainty.  As we move into an era of high-precision neutrino
astronomy, and considering the potential impact this will have on our
understanding of particle physics, it is imperative that we have a
precision determination of these weak currents.  Lattice QCD may be
the only rigorous method with which to determine these capture rates
with high precision unless a precise experimental determination is
made~\cite{Avignone:vh}.

In this work we explore the scattering states and bound states of the
two-nucleon sector at finite lattice volumes.  We first develop the
finite-volume effective field theory relevant to the very-low energy
interactions of two-nucleons, and we recover the exact eigenvalue
equation as well as several approximate formulas due to
L{\"u}scher~\cite{Luscher:1986pf,Luscher:1990ux}. We find several new
approximate formulas; we derive the leading finite-volume corrections
to the bound-state energy and we find perturbative formulas for the
lowest-lying energy levels for the case of a scattering length which
is large compared to the lattice size. Armed with this technology, we
consider simple unphysical limits of the scattering parameters to gain
some intuition, and then explore the two-nucleon sector itself.

\section{The Pionless Theory of NN Interactions in a Box}

\noindent 
An effective field theory (EFT) without pions, $\nopi$, has
been developed~\cite{Chen:1999tn,Kaplan:1998tg,Kaplan:1998we,vanKolck:1998bw} to
describe the very low-momentum interactions of two nucleons, with and without
electroweak probes. $\nopi$ exploits the sizable hierarchy between
the $S-$wave NN scattering lengths on the one hand, and the effective
ranges, shape parameters and pion Compton wavelength on the other, and
has been used successfully to perform relatively high-precision
calculations, some at the $\sim 1\%$ level.  Therefore, an EFT exists
that can be used to rigorously determine the behavior of low-momentum
nuclear observables in a finite volume, i.e. a lattice, and conversely
can be used to extract infinite-volume limits of lattice simulations of
two-nucleon observables. Furthermore, $\nopi$ has been used to
successfully compute the properties of three-nucleon
systems~\cite{Bedaque:2002mn}, and therefore calculations of
three-nucleon systems at finite volume should be possible as
well~\footnote{ As a point of interest,
it has been conjectured recently in
Ref.~\cite{Braaten} that QCD is very near the critical trajectory for
a renormalization-group limit cycle in the three-nucleon sector.}.

For NN scattering, the interaction between nucleons is described by a
series of local operators with an increasing number of derivatives
acting on the nucleon fields.  The scattering amplitude can be
computed in an elegant form using dimensional-regularization with
power-divergence subtraction (PDS)~\cite{Kaplan:1998tg,Kaplan:1998we}.
In this scheme, the coefficients of the operators in the Lagrange
density have natural size, even for unnaturally-large scattering
lengths.  
The scattering amplitude is identical to that of 
effective-range theory, and that found by solving the
Schr{\"o}dinger equation with a
pseudo-potential~\cite{vanKolck:1998bw}.
An important feature that
distinguishes $\nopi$ from other constructions is that electroweak interactions
can included systematically, in the same way they are included
in chiral perturbation theory ($\chi$PT).

\begin{figure}[!ht]
\vskip 0.15in
\centerline{{\epsfxsize=3.1in \epsfbox{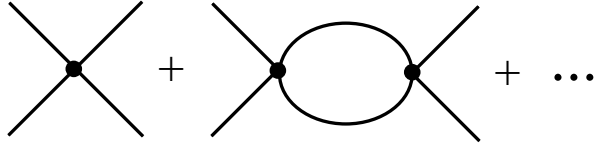}}}
\vskip 0.15in
\noindent
\caption{\it 
The diagrams in $\nopi$ which can be summed to give
the scattering amplitude. 
The small solid circle denotes an insertion of the infinite tower of
contact operators, $\sum C_{2n} (\mu)\ p^{2n}$.
}
\label{fig:bubble}
\vskip .1in
\end{figure}

In $\nopi$ (describing non-relativistic  baryons~\footnote{Relativistic corrections can
  be included in perturbation theory~\cite{Chen:1999tn}} each of mass $M$)
the exact two-body elastic scattering amplitude in the continuum
arises from the diagrams shown in Fig.~\ref{fig:bubble}, 
which can be resummed~\cite{Kaplan:1998tg,Kaplan:1998we} to give
\begin{eqnarray}
{\cal A} & = & { \sum C_{2n} (\mu)\ p^{2n}  \over
1 - I_0 \sum C_{2n} (\mu)\ p^{2n}}
\ \ \ \ ,\ \ \ \ 
I_0 \ = \ \left({\mu\over 2}\right)^{4-D} \int {d^{D-1}{\bf q}\over
  (2\pi)^{D-1}}
{1\over E-{|{\bf q}|^2\over M} + i \epsilon}
\ \ \ ,
\label{eq:fullsum}
\end{eqnarray}
where the $C_{2n} (\mu)\ $ are the renormalization-scale dependent
coefficients of operators with $2 n$ derivatives
acting on the nucleon fields (or equivalently with $n$ time derivatives),
$\mu$ is the dimensional-regularization scale, and $D$ is the number of space-time
dimensions. 
The loop integral $I_0$ is linearly divergent, and when defined with the PDS
subtraction scheme~\cite{Kaplan:1998tg,Kaplan:1998we} becomes
\begin{eqnarray}
I_0^{(PDS)} & = & - {M\over 4\pi}\ \left( \mu + i p \right)\ +\ {\cal O}(D-4)
\ \ \ ,
\end{eqnarray}
where $p=\sqrt{ M E}$ is the momentum of each nucleon in the center-of-mass,
and hence the scattering amplitude takes the usual form
\begin{eqnarray}
{\cal A} & = & {4\pi\over M}\ {1\over p\cot\delta - i p }
\ \ \ .
\end{eqnarray}
This unitary expression describes NN scattering below the onset of the
first inelastic threshold; that is, it is valid for $|{\bf p}|<\sqrt{m_\pi M}$,
where $m_\pi$ is the pion mass.  The subtraction-scale dependence of
the one-loop diagrams is exactly compensated by the corresponding
dependence of the coefficients $C_{2n} (\mu)$.  $\delta$ is the
energy-dependent $S-$wave phase shift (we will only consider $S-$wave
scattering but this construction generalizes to all partial waves).
It is clear from eq.~(\ref{eq:fullsum}) that $p\cot\delta$ is an
analytic function of $p^2$ for momenta less than the cut-off of
$\nopi$, which is $\sim m_\pi/2$.  We may therefore adopt the
effective-range expansion,
\begin{eqnarray}
p\cot\delta & = & -{1\over a} + {1\over 2} r_0\ p^2\ \sum_{i=0}^\infty\ (r_i^2\ p^2)^i
\ \ \ ,
\label{eq:er}
\end{eqnarray}
where $a$ is the scattering length, $r_0$ is the effective range, and the other
$r_i$ correspond to higher-order shape parameters~\footnote{We use the sign convention
for the scattering length that is traditionally used in nuclear physics. This is opposite
to the sign convention used by L{\"u}scher~\cite{Luscher:1986pf,Luscher:1990ux}}.

We are interested in the energy-eigenvalues of the NN system placed in
a box with sides of length $L$ with periodic boundary conditions.  
The scattering-state and bound-state  energy-eigenvalues
can be found by requiring 
the real part of the inverse scattering amplitude computed in the box to
vanish,
\begin{eqnarray}
{1\over \sum C_{2n}(\mu) \ p^{2n}}\ -\ {\rm Re}( I_{0}^{(PDS)}(L) ) & = & 0
\ \ \ ,
\label{eq:reamp}
\end{eqnarray}
where the infinite-volume integral in eq.~(\ref{eq:fullsum}) 
is replaced by a discrete sum over the momentum states allowed on the lattice,
\begin{eqnarray}
I_{0}(L)& = & 
{1\over L^3} \sum_{{\bf k}} {1\over E-{|{\bf k}|^2\over M} }
\ \ \ .
\end{eqnarray}
This discrete sum is also linearly divergent as the ultra-violet behavior of the
theory is unchanged, and its value in the PDS scheme is found by adding and
subtracting the corresponding infinite-volume integrals evaluated at $E=0$.
One of the infinite-volume integrals is evaluated with a momentum
cut-off, $|{\bf k}| \leq \Lambda$,
that is equal to the mode cut-off introduced to regulate the discrete sum, while
the other is evaluated with dimensional-regularization and PDS,
to give
\begin{eqnarray}
I_{0}^{(PDS)}(L)& = & 
-{M\over 4\pi} \mu\ +\ 
{1\over L^3} \sum_{{\bf k}}^\Lambda {1\over E-{|{\bf k}|^2\over M} }
\ +\ 
M \int^\Lambda { d^3 {\bf k}\over (2\pi)^3 }\ {1\over |{\bf k}|^2}
\ \ \ ,
\label{eq:pdssum}
\end{eqnarray}
and the limit  $\Lambda\rightarrow\infty$ is taken,
assuming that the lattice spacing vanishes. For any realistic simulation
there will be an upper bound on $\Lambda$, given by the edge of the first 
Brillouin zone~\cite{DavidL}.

\subsection{The Eigenvalue Equation}

\begin{figure}[!ht]
\vskip 0.28in
\centerline{{\epsfxsize=3.7in \epsfbox{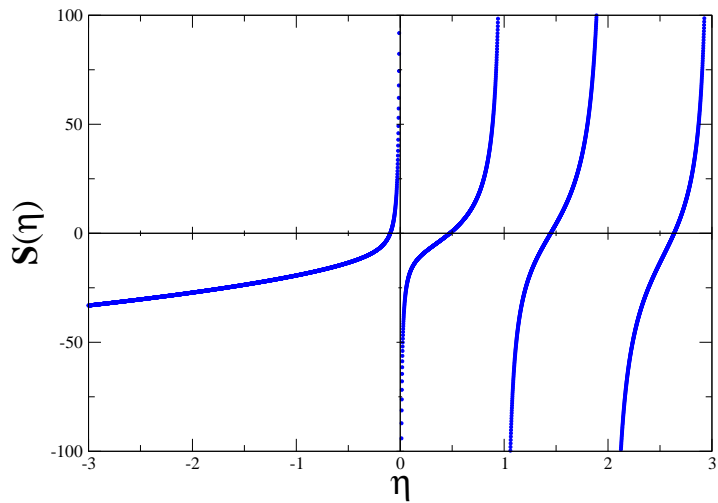}}} 
\vskip 0.15in
\noindent
\caption{\it 
A plot of ${\bf S}({\eta})$ vs.~${\eta}$ from eq.~(\protect\ref{eq:Sdefined}).
The function has poles for ${\eta}\geq 0$ and does not have poles for
${\eta}<0$.}
\label{fig:Paulo}
\end{figure}

\noindent 
The energies of the low-lying energy levels of
two-nucleons in a box with sides of length $L$ with periodic boundary
conditions~\cite{Luscher:1986pf,Luscher:1990ux,vanBaal:2000zc,Beane:2003yx}
and with their center-of-mass at rest
can now be determined in terms of $p\cot\delta$, from eq.~(\ref{eq:reamp})
and eq.~(\ref{eq:pdssum}).
Values of $p^2$ that solve~\footnote{
L{\"u}scher writes this expression
as~\cite{Luscher:1986pf,Luscher:1990ux}
\begin{eqnarray}
e^{2 i \delta_0(k)} & = & 
{{\cal Z}_{00}(1;q^2) + i\pi^{3/2} q\over {\cal Z}_{00}(1;q^2) - i\pi^{3/2} q}
\nonumber
\end{eqnarray}
where $q=p L/(2\pi)$. 
The three-dimensional zeta-functions are
\begin{eqnarray}
{\cal Z}_{00}(s;q^2)& = & 
{1\over\sqrt{4\pi}} \sum_{{\bf n}} \left({\bf n}^2-q^2\right)^{-s}
\nonumber
\end{eqnarray}
where the formally divergent functions are defined via analytic continuation. 
}
\begin{eqnarray}
p\cot\delta(p) \ =\ {1\over \pi L}\ {\bf S}\left(\,\left({Lp\over 2\pi}\right)^2\, \right)\ \ ,
\label{eq:energies}
\end{eqnarray}
with 
\begin{eqnarray}
{\bf S}\left(\,{\eta}\, \right)\ \equiv \ \sum_{{\bf j}}^{\Lambda_j} 
{1\over |{\bf j}|^2-{\eta}}\ -\  {4 \pi \Lambda_j}
\ \ \  ,
\label{eq:Sdefined}
\end{eqnarray}
give the location of all of the energy-eigenstates in the box, including
the bound states (with $p^2 < 0$).  The sum is over all three-vectors
of integers ${\bf j}$ such that $|{\bf j}| < \Lambda_j$ and where the
limit $\Lambda_j\rightarrow\infty$ is implicit (corresponding to the 
$\Lambda\rightarrow\infty$ limit in eq.~(\ref{eq:pdssum})).  
In Fig.~\ref{fig:Paulo} we plot ${\bf S}({\eta} )$ vs.~${\eta}$ from eq.~(\ref{eq:Sdefined}).
While our derivation of eq.~(\ref{eq:energies}) is valid
within the radius of convergence of $\nopi$, that is for $|{\bf p}|<m_\pi/2$,
we expect that eq.~(\ref{eq:energies}) remains valid as long as the energy of the two-nucleon 
states are below the pion-production threshold, $|{\bf p}|<\sqrt{m_\pi M}$~\cite{Luscher:1990ux}.
%

\subsection{Approximate Formulas}

\noindent There are two extreme limits that can be considered for the solution of
eq.~(\ref{eq:energies}).  First, there is the limit that L{\"u}scher
considers in his work in which $L\gg |a|$.  In this limit the solution
of eq.~(\ref{eq:energies}) smoothly approaches the infinite-volume limit.  
The energy of the two lowest-lying continuum states in
the $A_1$ representation of the cubic group~\cite{Mandula:ut} are
\begin{eqnarray}
E_0 & = & + {4\pi a\over M L^3}\left[\ 1\ -\ c_1 {a\over L}\ 
+\ c_2 \left({a\over L}\right)^2\ +\ ...\right]
\ +\ {\cal O}(L^{-6})
\ \ \ ,
\label{eq:e0}
\end{eqnarray}
where the coefficients are $c_1=-2.837297$, $c_2=+6.375183$, and 
\begin{eqnarray}
E_1 & = & {4\pi^2\over M L^2} - 
{12\tan\delta_0\over M L^2}\left[\ 
1 + c_1^\prime\tan\delta_0 + c_2^\prime \tan^2\delta_0\ +\ ...\ \right]
\ +\ {\cal O}(L^{-6})
\ \ \ ,
\label{eq:e1}
\end{eqnarray}
where $c_1^\prime=-0.061367 $,  $c_2^\prime=-0.354156$.
In addition, in the limit $L\gg a$, we have solved
eq.~(\ref{eq:energies})
for the location of the bound state that exists
for $a>0$ with an attractive interaction~\footnote{
The extension of the 
Chowla-Selberg formula to higher dimensions~\cite{Elizalde:1997jv} gives
\begin{eqnarray}
{\bf S}\left(-x^2 \right)\
&&\ \ 
\rightarrow\ \ 
- 2 \pi^2 x \ +\  6\pi e^{-2\pi x} + ...
\ \ \ ,
\end{eqnarray}
for large $x$,
where the ellipses denote terms exponentially suppressed by factors of
$e^{-4\pi x}$, or more.}
\begin{eqnarray}
E_{-1} & = & -{\gamma^2\over M}\left[\ 
1\ +\ {12\over \gamma L}\  {1\over 1-2\gamma (p\cot\delta)^\prime}\ 
e^{-\gamma L}\ +\ ...
\right]
\ \ \ ,
\label{eq:eb}
\end{eqnarray}
where $(p\cot\delta)^\prime={d\over dp^2}\ p\cot\delta$ evaluated at
$p^2=-\gamma^2$. The quantity $\gamma$ is the solution of
\begin{eqnarray}
\gamma\  +\  p\cot\delta |_{p^2=-\gamma^2} \ & = & 0
\ \ \ ,
\label{eq:pctdg}
\end{eqnarray}
which yields the bound-state binding energy in the infinite-volume limit.

In the limit where $L\ll |(p\cot\delta)^{-1}|$ (which is a useful limit to consider
when systems have unnaturally-large scattering lengths), 
the solution of eq.~(\ref{eq:energies})
gives the energy of the lowest-lying state to be
\begin{eqnarray}
\tilde E_0 & = & 
{4\pi^2\over M L^2}\left[\ d_1 \ +\   d_2\  L p\cot\delta_0\  + ...\
  \right]
\ \ \ ,
\label{eq:usE0}
\end{eqnarray}
where the coefficients are $d_1 = -0.095901$, $d_2 = +0.0253716$ and
where $p\cot\delta_0$ is evaluated at an energy $E={4\pi^2\over M L^2}\ d_1$.
The energy of the next level is 
\begin{eqnarray}
\tilde E_1 & = & 
{4\pi^2\over M L^2}\left[\  d_1^\prime \ +\  d_2^\prime \  L
  p\cot\delta_0
\  + ...\  \right]
\ \ \ ,
\label{eq:usE1}
\end{eqnarray}
where $d_1^\prime = +0.472895$, $d_2^\prime = +0.0790234$ and
where $p\cot\delta_0$ is evaluated at an energy $E={4\pi^2\over M L^2}\ d_1^\prime$.
The values of the $d_i^{(\prime)}$ are determined by zeroes of the 
three-dimensional zeta-functions, and 
the expressions for $E_i$ and $\tilde E_i$, excluding $E_{-1}$,
are valid for both-sign scattering lengths.

\subsection{A Toy Model : $a=\pm 1$ and $r_i=0$}

\noindent Let us consider an unphysical limit of NN scattering in
which the NN potential has zero-range but a scattering length of
$|a|=\pm 1$ in the infinite-volume limit.  Therefore, the scattering
amplitude is $p\cot\delta = -1/a$, as the effective range and all
shape parameters vanish, $r_i=0$.

\begin{figure}[!ht]
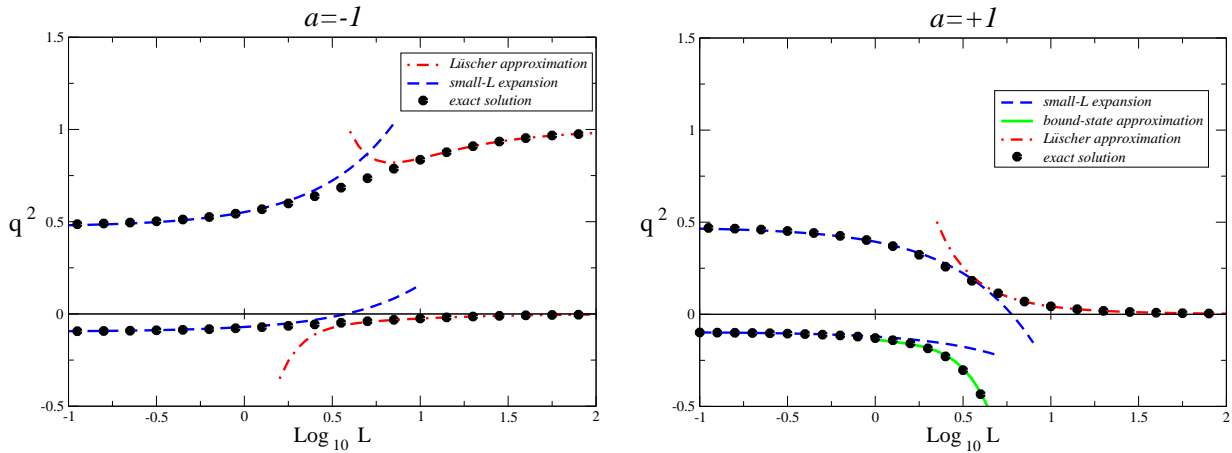

\vskip 0.15in
\centerline{{\epsfxsize=3.1in \epsfbox{TOYaminus.eps}}\hskip0.2in{\epsfxsize=3.1in \epsfbox{TOYaplus.eps}}} 
\vskip 0.15in
\noindent
\caption{\it 
The two lowest-lying solutions to eq.~(\protect\ref{eq:energies}) for 
$a=\pm 1~{\rm fm}$ and $r_i=0$.
The vertical axis is $q^2$, which is related to the energy by
$E=q^2 {4\pi^2 \over M L^2}$,
while the horizontal axis is $\log_{10} L$.
The left panel corresponds to $a=-1$ which can only arise from an attractive
potential.
The right panel  corresponds to $a=+1$ which can arise from both an attractive
and a repulsive potential.
For a repulsive potential the lower solution is absent.
The solid circles correspond to exact numerical solutions of
eq.~(\protect\ref{eq:energies}) .
The curves that match the exact solution at small $L$ result from 
eq.~(\protect\ref{eq:usE0}) and eq.~(\protect\ref{eq:usE1}),
while the curves that match the exact solution at large $L$ result from 
eq.~(\protect\ref{eq:e0}), eq.~(\protect\ref{eq:e1}),
and eq.~(\protect\ref{eq:eb}).
}
\label{fig:aplusminus}
\vskip .1in
\end{figure}
The system with $a=-1$ must result from an
attractive interaction, but one not attractive enough to yield a bound
state.  One might imagine that the potential is extremely attractive
and that the state with $a=-1$ is the one near threshold with many
other deep states present.  However, the deep states will be at the
cut-off of the theory, set by the range of the potential, and in the
limit we are considering these are infinitely deep.  The system with
$a=+1$ could result from a repulsive interaction in which case there
will be no bound state in this channel. However, $a=+1$ could also
result from an attractive interaction that is attractive enough to
give rise to a bound state near threshold. In the infinite-volume limit
of this second scenario, one must recover the
continuum of scattering states at positive energy and also the bound
state.

As $L\rightarrow 0$ the scattering lengths of opposite signs can be
identified with each other (as this is equivalent to taking the
$|a|\rightarrow\infty$ limit of the model) and so the levels become
degenerate.  One is tempted to think that the levels that are
degenerate in the $L\rightarrow 0$ limit are the same as those that
are degenerate in the $L\rightarrow\infty$ limit.  However, this
cannot be the case as the lowest state at $L=0$ in
the system with $a=+1$ smoothly becomes the bound state at $L=\infty$
(as a function of $L$) and a bound state does not exist in the $a=-1$
system. This can be seen clearly in Fig.~\ref{fig:aplusminus}.

It is apparent in both the $L\gg |a|$ and $L\ll |a|$
limits that the parameter arising in the asymptotic
expansion of the energies of the levels is $\sim 3 |a|/L$.
This can be seen clearly in Fig.~\ref{fig:aplusminus} where
L{\"u}scher's expressions in $a/L$ break down at $L\sim
3~{\rm fm}$ while the expressions in eq.~(\protect\ref{eq:usE0}) and
eq.~(\protect\ref{eq:usE1}), which are an expansion in $L/a$, remain
close to the exact solution out to $L\sim 3~{\rm fm}$.

\subsection{Low-energy NN Scattering in the S-wave}

\noindent 
It is well known that the scattering lengths and
effective ranges alone are sufficient to describe low-energy NN
scattering data to quite high precision.  This results in part from
the fact that the shape-parameters are much smaller than one would
naively guess. We therefore truncate the effective-range expansion
in our numerical analysis, and the 
power-counting of $\nopi$ dictates how the shape-parameters 
can be included in perturbation theory.  
At this order, the energy-levels of two-nucleons in the $\si$ channel 
whose center-of-mass is at rest
in a periodic box of size $L$ are found by solving
\begin{eqnarray}
&& {1\over a^{(\si)}} - {1\over 2} r^{(\si)}\  p^2
\ +\ 
{1\over \pi L} {\bf S}\left(\,\left({Lp\over 2\pi}\right)^2\, \right)
\ = \ 0
\ \ \ ,
\label{eq:exactE1S0}
\end{eqnarray}
for $p^2$, which is related to the energy of the NN system via $E=p^2/M$.
In the $\si$ channel the scattering length and effective range are
\begin{eqnarray}
a^{(\si)} & = & -23.714~{\rm fm}
\ \ ,\ \ 
r^{(\si)}\ =\ 2.734~{\rm fm}
\ \ .
\end{eqnarray}
\begin{figure}[!ht]
\vskip 0.15in
\centerline{{\epsfxsize=4.0in \epsfbox{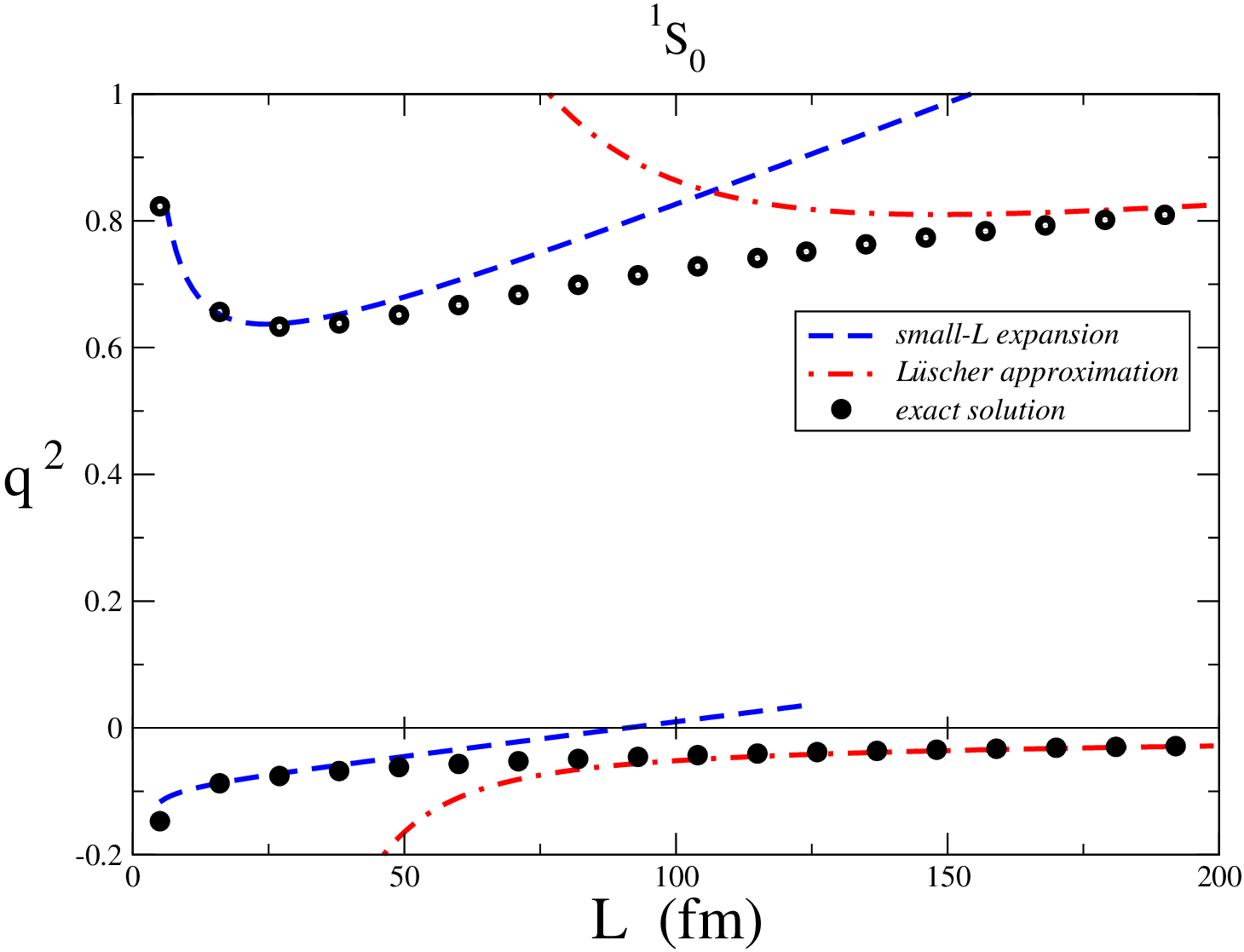}}} 
\vskip 0.15in
\noindent
\caption{\it 
The two lowest-lying energy-eigenstates of two nucleons in the $\si$ channel on
a lattice of size $L$.
The vertical axis is $q^2$,
where $E=q^2 {4\pi^2 \over M L^2}$,
while the horizontal axis is the lattice size $L$.
The solid circles correspond to the exact solution of
eq.~(\protect\ref{eq:exactE1S0}) .
The curves that are asymptotic to the exact solution at large $L$
correspond to L{\"u}scher's relations in
eq.~(\protect\ref{eq:e0}), eq.~(\protect\ref{eq:e1}),
while the curves that are asymptotic to the exact solution at 
smaller values of $L$
correspond to the expressions in 
eq.~(\protect\ref{eq:usE0}) and eq.~(\protect\ref{eq:usE1}).
}
\label{fig:E1S0}
\vskip .2in
\end{figure}
Despite the fact that the NN interaction is attractive in this channel
at long- and intermediate distances, there are no bound states of
$pp$, $np$ nor $nn$ in the $\si$ channel at the physical values of the
quark masses or $\Lambda_{\rm QCD}$~\footnote{One finds that
in all likelihood there is no bound state in this channel for quark
masses smaller than their physical values but it is possible that
bound states exists for quark masses somewhat larger than their
physical values~\cite{Beane:2002vq,Beane:2002xf,Epelbaum:2002gb}.  
This is an exciting possibility that can be explored
with lattice QCD.}.

One can see from Fig.~\ref{fig:E1S0} that L{\"u}scher's power-series
expansion of the energy-levels converges slowly in the $\si$
channel.  For the ground state one needs a box of size $L\gsim 80~{\rm
fm}\sim 3 a^{(\si)}$ while for the first excited state one needs a box
of size $L\gsim 150~{\rm fm}$ before these perturbative expansions converge to
the exact result.
Therefore, these asymptotic expressions will not be of
great utility to nuclear physicists 
in the near future. By contrast, the expressions we
have derived in the $L\rightarrow 0$ limit,
eq.~(\protect\ref{eq:usE0}) and eq.~(\protect\ref{eq:usE1}), are
applicable for boxes smaller than $L\lsim 50~{\rm fm}$.  However, the
crucial assumption, $L\gg r_i$, begins to break down for volumes with
$L\lsim 5~{\rm fm}$.  In table~\ref{table:levels} we show the
momenta of the first two states in the $\si$ channel.  
The lowest-level can be described by $\nopi$ on a
lattice with $L\sim 10~{\rm fm}$ but a lattice with $L\gsim 15~{\rm
fm}$ is required in order for $\nopi$ to describe the second state.
This is not to say that we cannot use the location of all the states
on a lattice with $L\sim 10~{\rm fm}$; for momenta outside
the range of validity of $\nopi$ and below pion-production threshold, i.e.
for $m_\pi/2<|{\bf p}|<\sqrt{m_\pi M}$, lattice results will have to be matched
directly to $p\cot\delta$ in the $\si$ channel as there is no effective-range expansion.
\begin{table}[ht]
\caption{Momenta of the lowest-lying levels of two-nucleons on the lattice.
An asterisk denotes momenta outside the range of validity of the effective-range 
expansion, $|{\bf p}|^{\rm max.}~=~m_\pi/2~\sim~70~{\rm MeV}$.
The energy of the state is $E=|{\bf p}|^2/M$, and the appearance of an ``i''
indicates a $-ve$ energy.
}
\label{table:levels}
\newcommand{\m}{\hphantom{$-$}}
\newcommand{\cc}[1]{\multicolumn{1}{c}{#1}}
\renewcommand{\tabcolsep}{0.2pc} 
\renewcommand{\arraystretch}{1.0} 
\begin{tabular}{ |@{}c | c | c | c | c | }
\hline
\multicolumn{1}{| c |}{}& 
\multicolumn{2}{| c |}{$\si$  \ \ $|{\bf p}|$ (MeV) } & 
\multicolumn{2}{| c |}{$\siii$ \ \ $|{\bf p}|$ (MeV)} \\
\hline
\ Lattice Size $L$ (fm) \ & 
\ 1st \  & 
\ 2nd  \ & 
\ Deuteron \  & 
\ 1st \  
\\
\hline
\hline
$1000$ & 0.1 i & 1.21 & 45.5 i & 0.052\\
\hline
$100$ & 2.6 i & 10.52 & 45.5 i & 1.76\\
\hline
$25$ & 13.8 i & 39.3 & 45.8 i & 18.25\\
\hline
$15$ & 24.6  i & 67.0 & 49.9 i & 44.61\\
\hline
$10$ & 39.0  i & 104.3 (*) & 61.3 i & 83.1 (*)\\
\hline
$5$ & \ 94.4  i (*)\   &\  224.7 (*)\  &\  116.5  i (*)\ &\  206.5 (*)\ \\
\hline
\end{tabular}
\end{table}

The $\siii-\diii$ channel is somewhat complicated by the fact that the
tensor interaction gives rise to mixing between the $\siii$ and the
$\diii$ channels.  On the lattice this means that different
representations of the cubic group will mix due to the tensor
component of the NN interaction.  In particular, the $A_1$
representation will mix with the $E$ and $T_2$ representations~\cite{Mandula:ut}.
However, the power-counting of $\nopi$ dictates that we can ignore
contributions from this mixing and the $\diii$ channel at the order to
which we are working~\cite{Chen:1999tn}.  Therefore, at this order,
the energy levels of two nucleons in the $\siii$ channel can be found
by solving
\begin{eqnarray}
&& {1\over a^{(\siii)}} - {1\over 2} r^{(\siii)}\  p^2
\ +\ 
{1\over \pi L} {\bf S}\left(\,\left({Lp\over 2\pi}\right)^2\, \right)
\ = \ 0
\ \ \ .
\label{eq:exactE3S1}
\end{eqnarray}
In the $\siii$ channel the effective-range parameters are
\begin{eqnarray}
a^{(\siii)} & = & +5.425~{\rm fm}
\ \ ,\ \ 
r^{(\siii)}\ =\ 1.75~{\rm fm}
\ \ .
\end{eqnarray}

\begin{figure}[!ht]
\vskip 0.15in
\centerline{{\epsfxsize=4.0in \epsfbox{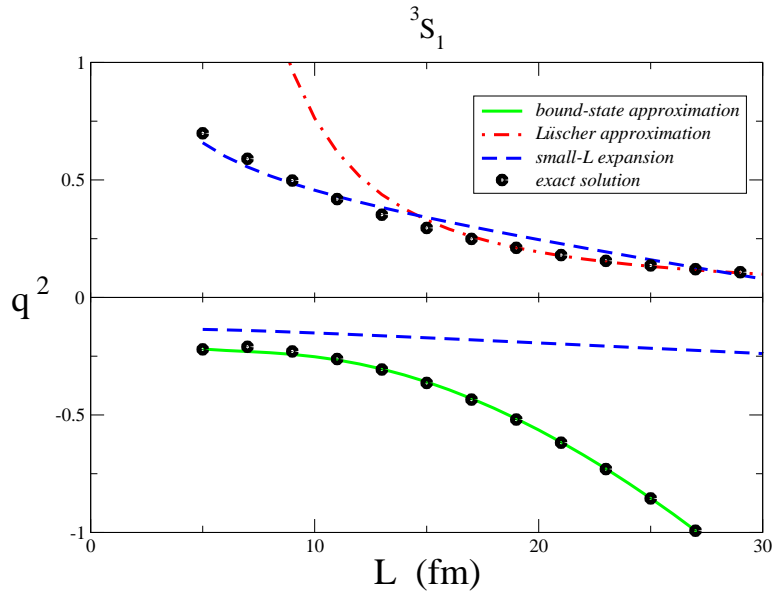}}} 
\vskip 0.15in
\noindent
\caption{\it 
The two lowest-lying energy-eigenstates of two nucleons in the $\siii$ channel on
a lattice of size $L$.
The vertical axis is $q^2$,
where $E=q^2 {4\pi^2 \over M L^2}$,
while the horizontal axis is $L$.
The solid circles correspond to  exact solutions of
eq.~(\protect\ref{eq:exactE3S1}) .
The curves that are asymptotic to the exact solution at large $L$
correspond to L{\"u}scher's relation in
eq.~(\protect\ref{eq:e0}) and the relation for the bound-state energy, 
eq.~(\protect\ref{eq:eb}),
while the curves that are asymptotic to the exact solution at small $L$
correspond to the expressions in 
eq.~(\protect\ref{eq:usE0}) and eq.~(\protect\ref{eq:usE1}).
}
\label{fig:E3S1}
\vskip .2in
\end{figure}
One can see from Fig.~\ref{fig:E3S1} that L{\"u}scher's power-series
expansion of the energy-levels converges to the exact solution for the
lowest-lying continuum state when $L\gsim 15~{\rm fm}\sim 3
a^{(\siii)}$.  
However, our expression for the deuteron binding energy at 
finite-volume appears to work even at  significantly smaller volumes.
Unfortunately, the expressions that we have derived for small volumes do not
converge well to the exact solution in this channel.  This is because the
scattering length is only  a factor of $\sim 3$ larger than the effective range
and thus the expansion parameter $L p\cot\delta$ is not small enough over the
entire range of $L$. Perhaps higher-order contributions will improve the agreement.
The deuteron state can be described with $\nopi$ for lattices
with $L\gsim 10~{\rm fm}$, but the lowest-lying continuum state can be
described by $\nopi$ only for $L\gsim 15~{\rm fm}$. In table~\ref{table:levels} we show 
the momenta of the deuteron bound state and the lowest-lying scattering state 
in the $\siii$ channel. 

Beyond the range of validity of $\nopi$, the $\siii$ channel is significantly different
than the $\si$ channel. As mentioned above, in $\nopi$ $\siii-\diii$ mixing is subleading
in the expansion. However in the pionful theory, i.e. for $m_\pi/2<|{\bf p}|<\sqrt{m_\pi M}$,
$\siii-\diii$ mixing appears at leading order in the EFT expansion~\cite{Beane:2001bc}.
Therefore, in this range of energies eq.~(\ref{eq:exactE3S1}) is {\it not} valid
and one must solve a coupled system of integral equations.

\section{Conclusions}

\noindent Lattice QCD calculations of the scattering lengths and
effective ranges in the two-nucleon sector would be a significant
milestone toward rigorous calculations of nuclear properties and
decays. Aside from providing essential information about the
quark-mass dependence of nuclear physics, such calculations by
themselves will not significantly improve our ability to compute
nuclear properties, as the scattering amplitudes are already
well-known experimentally.  However, we would be in a position to
compute electroweak matrix elements between two-nucleon states, and
thereby provide information that will be vital to future calculations
of electroweak processes involving nuclei.  Knowledge of these
processes will directly affect analysis of data from present and
future neutrino observatories.

Exact solutions to L{\"u}scher's general formula for the energy-levels
of the two-nucleon system on a lattice with periodic boundary
conditions will allow for the extraction of scattering parameters from
simulations with lattice volumes that are much smaller than naively
estimated.  It would appear that simulations on lattices with $L\gsim
15~{\rm fm}$ will make it possible to extract both the scattering
lengths and effective ranges in the two-nucleon sector in a
straightforward way.  This is contrary to the expectation that
lattices with $L\gg |a|$ are required to determine the scattering
parameters.  The extraction of useful information from simulations
with lattices $L\lsim 10~{\rm fm}$ will require direct matching to
$p\cot\delta$ in the spin singlet channel, as the pionless theory and
thus the effective-range expansion, will no longer be appropriate. In
the spin-triplet channel there remains the additional challenge of
formulating the finite-volume eigenvalue equations which account for
mixing between $S-$ and $D-$waves. These finite-volume calculations
are a vital component of the technology required to make rigorous
statements about nuclear processes directly from QCD.

\acknowledgments

\noindent 
We would like to thank Will Detmold, 
David Kaplan, David Lin and Steve Sharpe for
helpful discussions and Andrei Starinets for referring us to useful
literature on multi-variate sums. The work of SRB was partly supported
by DOE contract DE-AC05-84ER40150, under which the Southeastern
Universities Research Association (SURA) operates the Thomas Jefferson
National Accelerator Facility. PFB was supported by the Director,
Office of Energy Research, Office of High Energy and Nuclear Physics,
and by the Office of Basic Energy Sciences, Division of Nuclear
Sciences, of the U.S.~Department of Energy under Contract
No.~DE-AC03-76SF00098.  MJS is supported in part by the U.S.~Dept. of
Energy under Grant No.~DE-FG03-97ER4014.  AP is supported by the MCyT
under Grant No.~DGICYT BFM2002--01868 and by the Generalitat de
Catalunya under Grant No.~SGR2001--64.

\end{document}